\documentclass[fleqn,usenatbib,useAMS]{mnras}

\usepackage{graphicx}	% Including figure files
\usepackage{mwe}
\usepackage{amsmath}	% Advanced maths commands
\usepackage{amssymb}	% Extra maths symbols
\usepackage{multicol}        % Multi-column entries in tables
\usepackage{bm}		% Bold maths symbols, including upright Greek
\usepackage{pdflscape}	% Landscape pages
\usepackage{multirow}
\usepackage{ulem}
\usepackage{comment}
\usepackage{array}
\usepackage{caption}
\usepackage{subcaption}
\usepackage{lineno}
\usepackage{float}
\usepackage{newtxtext,newtxmath}

\includecomment{comment} %show comments
\usepackage[T1]{fontenc}
\usepackage{ae,aecompl}
\usepackage{newtxtext,newtxmath}
\usepackage{caption}
\usepackage{float}

\title[Supernovae from 25 M$_{\odot}$ Pop III star]{Evolution of Rotating 25\,M$_{\odot}$ Population III star: Physical Properties and Resulting Supernovae}

\author[Aryan et al]{Amar Aryan$^{1,2}$\thanks{Contact e-mail: \href{mailto:amararyan941@gmail.com}{amararyan941@gmail.com}
\href{mailto:amar@aries.res.in}{amar@aries.res.in}},
Shashi Bhushan Pandey$^{1}$, Rahul Gupta$^{1,2}$, and Amit Kumar Ror$^{1}$
\\
$^{1}$Aryabhatta Research Institute of Observational Sciences (ARIES), Manora Peak, Nainital-263002, India \\
$^{2}$Department of Physics, Deen Dayal Upadhyaya Gorakhpur University, Gorakhpur-273009, India}
\date{}

\pubyear{}
\begin{document}
\label{firstpage}
\pagerange{\pageref{firstpage}--\pageref{lastpage}}
\maketitle

\begin{abstract}
In this Letter, we report the outcomes of 1-D modelling of a rotating 25\,M$_{\odot}$ zero-age main-sequence Population III star up to the stage of the onset of core collapse. Rapidly rotating models display violent and sporadic mass losses after the Main-Sequence stage. In comparison to the solar metallicity model, Pop III models show very small pre-supernova radii. Further, with models at the stage of the onset of core collapse, we simulate the hydrodynamic simulations of resulting supernovae. Depending upon the mass losses due to corresponding rotations and stellar winds, the resulting supernovae span a class from weak Type II to Type Ib/c. We find that the absolute magnitudes of the core-collapse supernovae resulting from Pop III stars are much fainter than that resulting from a solar metallicity star. From our simulation results, we also conclude that within the considered limits of explosion energies and Nickel masses, these transient events are very faint, making it difficult for them to be detected at high redshifts.
\end{abstract}

%%%%%%%%%%%%%%%%%%%%%%%%%%%%%%%%%%%%%%%%%%%%%%%%%%
%%%%%%%%%%%%%%%%% BODY OF PAPER %%%%%%%%%%%%%%%%%%
\section{Introduction}
\label{sec:intro}
The first generations of stars formed out of uncontaminated matter, initially comprised only of the first two stable elements, Hydrogen (H) and Helium (He) from the periodic table, are considered as the population III (Pop III) stars. Due to the insufficiency of coolants in the primordial gas, it is hypothesised that the Pop III stars were massive intrinsically \citep[][]{1983MNRAS.205..705S,1997ApJ...474....1T,1999ApJ...527L...5B,  2001ApJ...548...19N, 2002Sci...295...93A, 2007ApJ...661...10B, 2010MNRAS.401L...5S, 2015MNRAS.448..568H}. However, there have been multiple studies to apprehend the possibility of the existence of Pop III stars having low masses. In recent simulations, it has been found that the formation of pristine, metal-free stars at low to intermediate masses could potentially be due to the fragmented accretion disks around massive Pop III protostars \citep[][]{2009Sci...325..601T,2010MNRAS.403...45S,2011Sci...334.1250H,2011Sci...331.1040C,2012MNRAS.424..399G,2014ApJ...781...60H, 2015MNRAS.448..568H,2016MNRAS.462.1307S,  2018MNRAS.479..667R, 2020MNRAS.494.1871W}. 
%While performing population synthesis simulations, \citet[][]{2014MNRAS.442.2963K} have explored the possibility of the indirect confirmation regarding the existence of the massive Pop III stars utilising the gravitational waves. In their work, the authors have concluded that the gravitational waves generated due to the inspiraling of the two companions can be detected up to z $=$ 0.28 by a number of gravitational wave detectors including KAGRA,  Advanced LIGO,  Advanced Virgo, and GEO network, if the characteristic mass of the Pop III binary black holes is around 30\,M$_{\odot}$.
On the-low mass Pop III stars, \citet[][]{2016ApJ...826....9I} have found that the Pop III stars having masses $<$\,0.8\,M$_{\odot}$ would have longer lifetimes as compared to the cosmic time, therefore such low-mass stars could linger around to be detected in our Milky Way itself. 

Further, Pop III stars were responsible for the enrichment of the early universe by spreading metals heavier than He through violent supernova (SN) explosions or possibly through sporadic mass losses due to vigorous stellar winds \citep[][]{2000MNRAS.319..539F,2001ApJ...557..126A}. 
The study by \citet[][]{2018MNRAS.475.4378C} shows that a core-collapse supernova (CCSN) from a Pop III star could cause the minihalo to undergo internal-enrichment. This causes the metallicity to be -5 $\lesssim$\,[Fe/H]\,$\lesssim$-3 in the recollapsing region. Thus, internal-enrichment caused by a CCSN from a Pop III star can explain the stars which are extremely metal-poor.
In a relatively recent work, the authors of \citet[][]{2020MNRAS.491.4387K} have estimated the dose of heavy elements introduced by massive Pop III stars. In doing so, they considered the amount of heavy elements synthesised only from pair-instability supernovae (PISNe) or core-collapse supernovae (CCSNe) explosions of massive Pop III stars. They found that the heavy elements introduced by Pop III stars are usually much more than those from galaxies found in the low-density regions. Besides the above-mentioned studies involving Pop III stars, there have also been investigations to recognize the influence of Pop III stars on cosmic reionization \citep[e.g.,][]{1997ApJ...476..458H,2000ApJ...528L..65T,2001PhR...349..125B,2001MNRAS.324..381C, 2013RPPh...76k2901B}, and dust formation \citep[][]{2001MNRAS.325..726T}.

There have been multiple studies to understand the evolution of the Pop III stars. \citet[][]{2003A&A...399..617M} and \citet[][]{2008A&A...489..685E} have studied the evolution of Pop III stars by assuming solid-body and differential rotation, respectively. \citet[][]{2010ApJ...724..341H} have discussed the nucleosynthesis and evolutions of non-rotating Pop III stars. They also generated the light curves of the resulting transients from the non-rotating models corresponding to different explosion energies. \citet[][]{2012A&A...542A.113Y} have discussed the evolution of massive Pop III stars having masses in the range of [10 -- 1000]\,M$_{\odot}$ and have investigated the consequences of including rotation and magnetic fields. Due to the chemically homogeneous evolution, the rapidly rotating high-mass stars could result into a class of energetic transients including type Ib/c supernovae (SNe), gamma-ray bursts, hypernovae, and PISNe. In their work, \citet[][]{2012A&A...542A.113Y} have prepared a phase diagram in the plane of mass and rotational velocity at zero-age main-sequence (ZAMS) and discussed the culminating fates of Pop III stars. The authors of \citet[][]{2018ApJS..234...41W} have investigated the evolutions of non-rotating Pop III stars in the mass range of [1 -- 1000]\,M$_{\odot}$ considering no mass loss and have also discussed chances of the observability  of an individual Pop III star. In a recent work, \citet[][]{2021MNRAS.501.2745M} have studied a grid of Pop III models having masses in the range of [1.7 -- 120]\,M$_{\odot}$ and explored the effect of changing the initial rotational velocity from 0 to 40 per cent of critical rotational velocity.

Taking such studies one step further, in this work, we study the entire evolution (from ZAMS up to the stage of the onset of core collapse) of a 25\,M$_{\odot}$ Pop III star and investigate the effect of rotation on the final fates. %{The work in this study has been significantly supported by \citet[][]{2012A&A...542A.113Y}.}
Following the phase diagram in \citet[][]{2012A&A...542A.113Y}, the resulting supernovae (SNe) from a 25\,M$_{\odot}$ ZAMS star will either be weak Type II or Type Ib/c depending upon the initial rotations. For the first time in this work, we have evolved the rotating and non-rotating Pop III models together up to the stage of the onset of core collapse and further performed the hydrodynamic simulations of their synthetic explosions showing the light curves of resulting transients.

We have divided the entire Letter into four sections. After providing a brief introduction of literature in Section~\ref{sec:intro}, we discuss the numerical setups and physical properties of the models in Section~\ref{sec:mesa}. The numerical setups to simulate the hydrodynamic explosion of models are discussed in Section~\ref{sec:snec}. Finally, the major outcomes from the entire evolutions of the models along with their synthetic explosions are discussed in Section~\ref{sec:results and discussion}. In this Section, we also provide the implications and discussions of the simulation results presented in the underlying work.

\section{Stellar evolution using {\tt MESA}}
\label{sec:mesa}
\subsection{Numerical setups}
We employ one of the state-of-the-art and 1-D stellar evolution codes, {\tt MESA} to perform the stellar evolutions of 25\,M$_{\odot}$ ZAMS stars with zero metallicity (Z $=$ 0.00) and different initial rotations. In this simulation work, we have utilised the {\tt MESA} version r22.05.1 \citep[][]{2011ApJS..192....3P, 2013ApJS..208....4P, 2015ApJS..220...15P, 2018ApJS..234...34P}. We start with a non-rotating, zero metallicity model and increase the angular rotational velocity ($\Omega$) in units of 0.2 times the critical angular rotational velocity ($\Omega_{\rm crit}$) up to $\Omega$ / $\Omega_{\rm crit}$ $=$ 0.8. Following \citet[][]{2013ApJS..208....4P}, the critical angular rotational velocity is expressed as $\Omega_{\rm crit}^2 = (1-L/L_{\rm edd})GM/R^3$, where $L_{\rm edd}$ is the Eddington luminosity. As specified in the default {\tt MESA} setups, when the ratio of the luminosity from nuclear processes and the overall luminosity of the model at a particular stage reaches 0.4 (set by {\tt Lnuc\_div\_L\_zams\_limit} = 0.4 in {\tt MESA}), the model is assumed to have reached ZAMS. Adopting the Ledoux criterion, convection is modelled utilising the mixing theory of \citet[][]{1965ApJ...142..841H}. We have adopted a mixing-length-theory parameter ($\alpha_{\rm MLT}$) $=$ 2.0 in the whole analysis. Following \citet[][]{1985A&A...145..179L}, semi-convection is modelled by setting an efficiency parameter of $\alpha_{\mathrm{sc}}$ = 0.01.  To model the thermohaline mixing, we follow \citet[][]{1980A&A....91..175K} by fixing the efficiency parameter to $\alpha_{\mathrm{th}}$ = 2.0. In order to model the convective overshooting, we use the diffusive approach as presented in \citet[][]{2000A&A...360..952H} and set $f = 0.004$ and $f_{0} = 0.001$ for all convective cores and shells. The stellar winds from rotating Pop III stars are also theorised to make contributions to the enrichment of the early universe with heavy metals, so, to model the stellar winds of the Pop III stars in our study, we use the 'Dutch' wind scheme \citep[][]{2009A&A...497..255G} and set the scaling factor ($\eta$) to 0.5. This wind scheme incorporates the outcomes from multiple works for different situations; With m$_{\rm H}$ representing the surface mass fraction of Hydrogen, (a) when the effective temperature, T$_{\rm eff} > 10^4$\,K along with m$_{\rm H}$ being greater than 0.4, the outcomes of \citet[][]{2001A&A...369..574V} are used; (b) when T$_{\rm eff} > 10^4$\,K combined with the m$_{\rm H}$ being lesser than 0.4, the results of \citet[][]{2000A&A...360..227N} are used; and finally (c) the wind scheme presented in \citet[][]{1988A&AS...72..259D} is used in {\tt MESA} for the condition with T$_{\rm eff} < 10^4$\,K. 

Starting from the pre-ZAMS, the 1-D stellar evolutions of the Pop III models are performed till they reach the stage of the onset of the Iron core collapse. The onset of core collapse is marked when the infall velocity of the Iron core exceeds the specified Iron core infall velocity limit of 100\,km\,s$^{-1}$ (set by {\tt fe\_core\_infall\_limit = 1d7} in {\tt MESA}). In this work, the models have been named in a way that they contain the pieces of information including the ZAMS mass, metallicity (Z), and rotation ($\Omega$ / $\Omega_{\rm crit}$ indicated by "Rot"). For example, the model M25\_Z0.00\_Rot0.0 indicates a 25\,M$_{\odot}$ ZAMS star with Z = 0.00, and $\Omega$ / $\Omega_{\rm crit}$ = 0.0. For comparison purposes, we have also performed the evolution of a solar metallicity (Z = 0.02) model with the same ZAMS mass of 25\,M$_{\odot}$.

\begin{figure}	
    \includegraphics[width=\columnwidth]{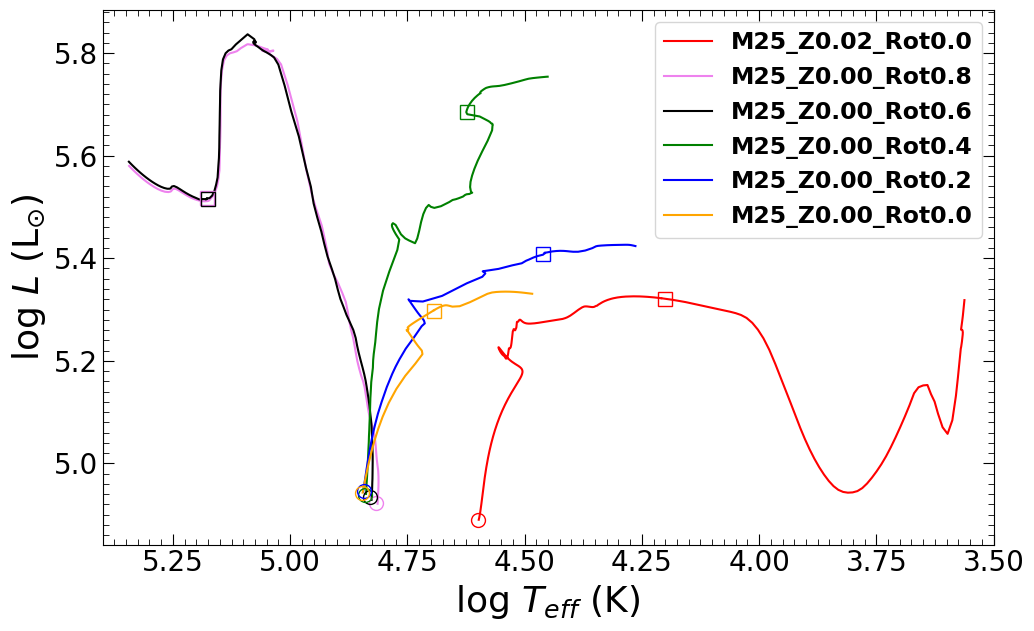}
    \caption{Evolution of the Pop III models having different rotations on the HR diagram. The arrivals of the models on ZAMS are shown by hollow circles while the core-He exhaustion stages of the models have been marked by hollow squares. The solar metallicity (Z = 0.02) model evolutionary track has also been shown for comparison.}
    \label{fig:HR_Diagram}
\end{figure}

\begin{figure}	
    \includegraphics[width=\columnwidth]{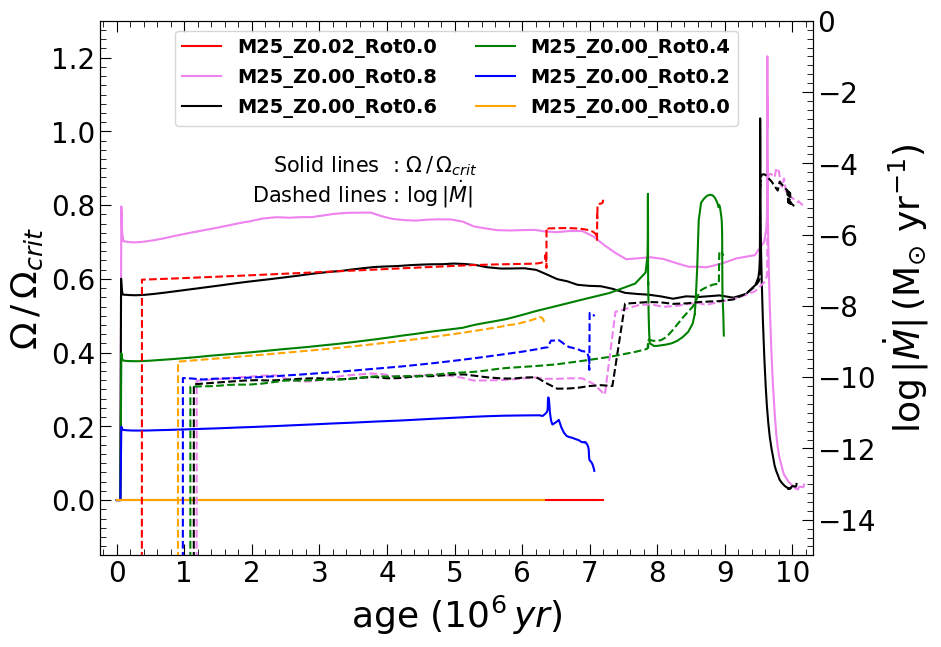}
    \caption{ The evolution of angular rotational velocity ($\Omega$) in units of critical angular rotational velocity ($\Omega_{\rm crit}$) along with corresponding mass loss rate (log |$\dot{M}$|) evolution.}
    \label{fig:Omega_plot}
\end{figure}

\subsection{Physical Properties of the Models}
%\subsubsection{Evolution on HR diagram}
Figure~\ref{fig:HR_Diagram} displays the evolutions of Pop III models with 25\,M$_{\odot}$ ZAMS mass each, on the Hertzsprung–Russell (HR) diagram along with a similar ZAMS mass model having solar metallicity (Z = 0.02) for comparison purposes.  Compared to a solar metallicity model, the Pop III models reach the ZAMS at higher effective temperatures ($T_{\rm eff}$) but show nearly similar ZAMS luminosities. Thus, the Pop III models are bluer than the solar metallicity model.
%Further, the Pop III models show slightly different pre-ZAMS evolution compared to a solar metallicity model. 
Among Pop III models, at ZAMS, the models with higher initial angular rotational velocities possess lower luminosities and lower effective temperatures; a well-known effect of rotation as mentioned in \citet[][]{2008A&A...489..685E}. It is also evident from this figure that beyond ZAMS, the models with higher initial angular rotational velocities possess higher luminosities and higher effective temperatures as well for most of their evolutionary paths. Similar results were also obtained in the case of \citet[][]{2012A&A...542A.113Y}.

\begin{figure}
\includegraphics[width=\columnwidth,angle=0]{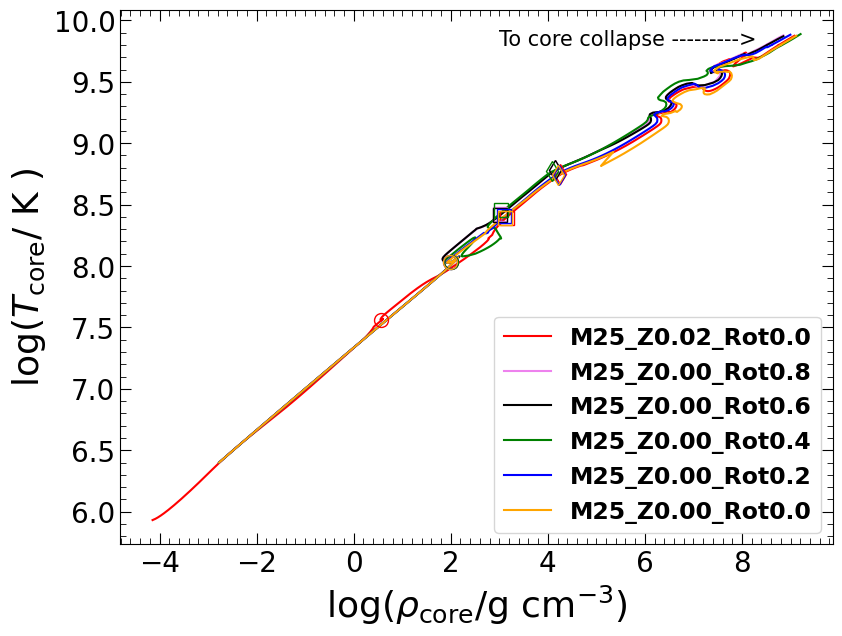}
   \caption{The variations of core-temperature ($T_{\rm core}$) vs core-density ($\rho_{\rm core}$) curves throughout the course of evolution of the models on the HR Diagram. The arrival on the ZAMS, exhaustion of core-He burning, and exhaustion of core-C burning have been marked by hollow circles, squares, and diamonds, respectively.}
    \label{fig:density_temp}
\end{figure}
%\subsubsection{Rotations and Mass losses}
Figure~\ref{fig:Omega_plot} shows the variations of $\Omega$ / $\Omega_{\rm crit}$ and corresponding mass loss rates (log |$\dot{M}$|) as the models evolve up to the stage of the onset of core collapse. Initially, the models touch the specified $\Omega$ / $\Omega_{\rm crit}$ values and then settle to new $\Omega$ / $\Omega_{\rm crit}$ values as they evolve further. This shows that a perturbation imposed on an equilibrium model experiences a transient response before the system settles into a new equilibrium configuration. During the last evolutionary stages, the rapidly rotating models (with $\Omega$ / $\Omega_{\rm crit}$ = 0.6 and 0.8) show chaotic rotations exceeding the critical rotational velocities which are responsible for the dynamic events to occur as indicated by corresponding heavy mass loss rates during these phases. Our rapidly rotating massive Pop III stars dredge up a large amount of CNO elements up to the surface during the core-He burning stage. It dramatically increases the surface metallicity, which eventually strongly boosts the radiative mass loss through a mechanism similar to that discussed in \citet[][]{2007A&A...461..571H}. Thus, the rapidly rotating models are significantly stripped compared to the slow- or non-rotating models. We have also shown the Kippenhahn diagram for one slow-rotating model (M25\_Z0.00\_Rot0.2) and one rapidly rotating model (M25\_Z0.00\_Rot0.8) in Figure~\ref{fig:kippenhahn}.
A few more important physical properties including radii and effective temperatures ($T_{\rm eff}$) at various stages are listed in Table~\ref{tab:MESA_MODELS}.

%\subsubsection{Pre-SN radii and the evolution of $\rho_{\rm core}$ vs $T_{\rm core}$}

The overall variations of the core-density ($\rho_{\rm core}$) vs core-temperature ($T_{\rm core}$) curves for the entire evolutions of the models up to the onset of core collapse are shown in Figure~\ref{fig:density_temp}. The arrival of the models on ZAMS, the exhaustion of core-He burning phases, and the exhaustion of core-C burning stages have been indicated by the hollow circles, squares, and diamonds, respectively. Compared to a solar metallicity model, the Pop III models ignite the H-burning in their respective cores at higher $\rho_{\rm core}$ and $T_{\rm core}$, which is due to the lack of CNO elements in Pop III stars needed to ignite the CNO cycle \citep[][]{2008A&A...489..685E}. During the last evolutionary stages, all the models have exceeded the core temperatures of $\sim$10$^{9.9}$\,K; the perfect condition for the cores to collapse under their own gravity. Thus, the models have now reached the stage of the onset of core collapse.

%\begin{figure*}
%  \centering
%  \includegraphics[height=9cm,width=\textwidth]{lc.png}
%    \caption{The U, B, V, R, and I band light curves resulting from the synthetic explosions of Pop III models using {\tt SNEC}. The non-rotating and slowly rotating models ($\Omega\,\leq\,0.4\,\Omega_{\rm crit}$) form a class of weak Type II SNe, while the rapidly rotating models ($\Omega\,\ge\,0.6\,\Omega_{\rm crit}$) result into Type Ib/c SNe. The light curves resulting from the non-rotating, solar metallicity model are also shown for comparison.} 
%    \label{fig:lc}
%\end{figure*}
\begin{figure*}
\centering
    \includegraphics[height=6.8cm,width=8.8cm,angle=0]{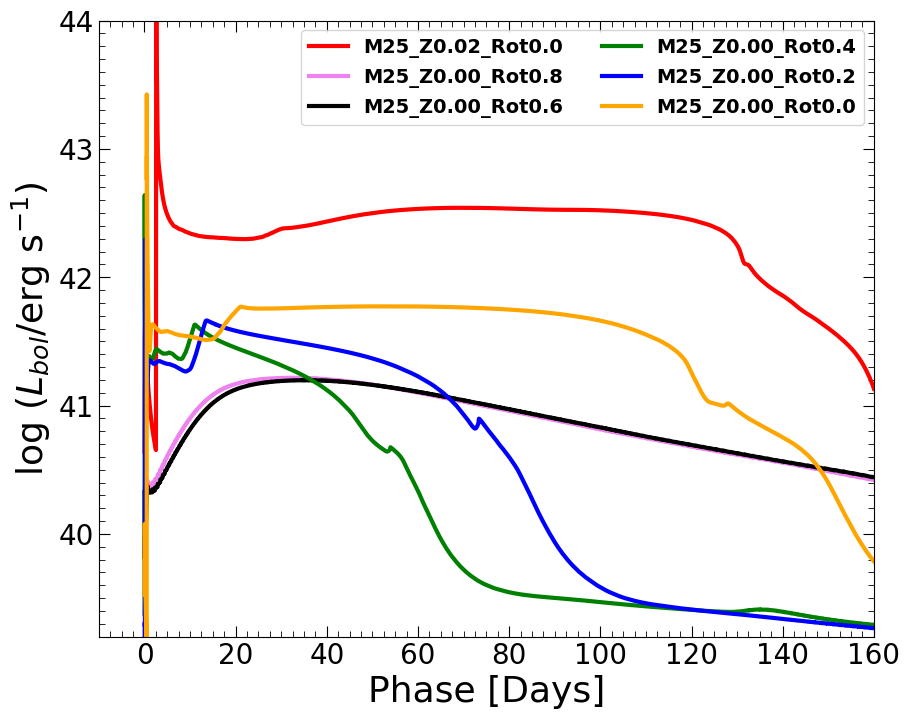}
    \includegraphics[height=6.8cm,width=8.8cm,angle=0]{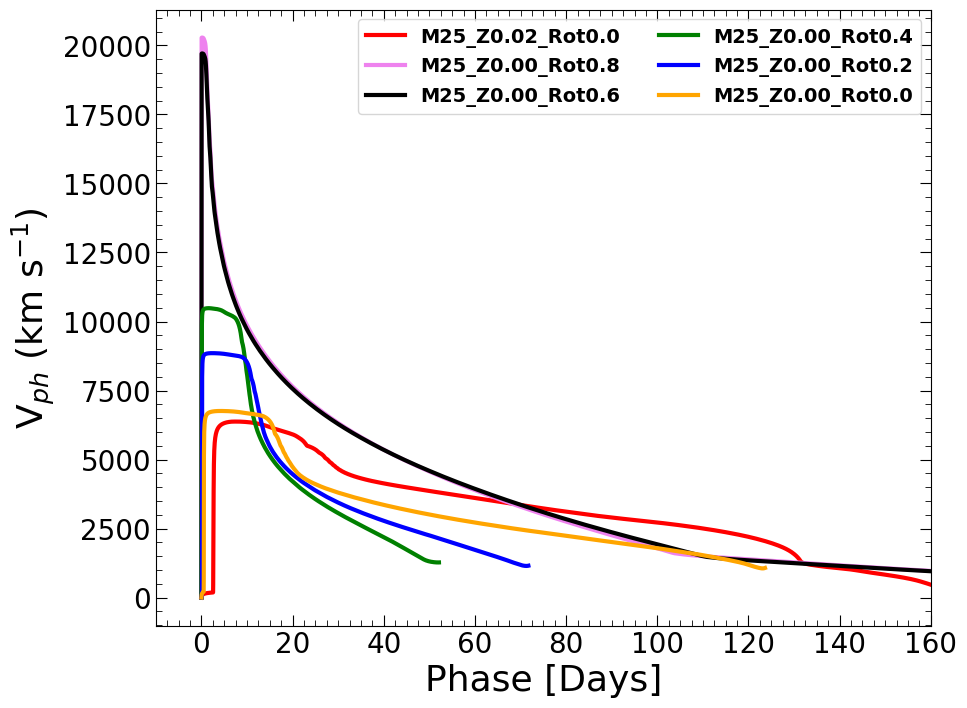}
   \caption {{\em Left:} The bolometric luminosity light curves resulting from the synthetic explosions of models using {\tt SNEC}. {\em Right:}  Corresponding photospheric velocity evolutions. Results of the non-rotating, solar metallicity model are also shown for comparison.}
    \label{fig:blc_vel}
\end{figure*}

\begin{table*}
\caption{The ZAMS and pre-SN properties of the Pop III models using {\tt MESA} along with the {\tt SNEC} explosion parameters. }
\label{tab:MESA_MODELS}
\begin{center}
{\scriptsize
\begin{tabular}{ccccccccccccc}
\hline \hline

 & &  ZAMS & & \hspace{1.3cm}\vline&  &  &  Pre-SN &\hspace{1.3cm}\vline & &  Explosion\\
\hline
Model Name	& $M_{\rm ZAMS}^{a}$	& $T_{\mathrm{eff}}$  & $R_{\mathrm{ZAMS}}^{b}$ & $L_{\rm ZAMS}^{c}$  &	$M_{\rm Pre-SN}^{d}$	& $T_{\mathrm{eff}}$  & $R_{\mathrm{Pre-SN}}^{e}$ & $L_{\rm pre-SN}^{f}$ 	&	$M_{\mathrm{c}}^{g}$ & $M_{\mathrm{ej}}^{h}$ & $M_{\mathrm{Ni}}^{i}$ &	$E_{\mathrm{exp}}^{j}$ 	\\
	&	(M$_{\odot}$) & K	&	(R$_{\odot}$)	 &  (L$_{\odot}$) &      (M$_{\odot}$) & K	&	(R$_{\odot}$)	 &  (L$_{\odot}$) & (M$_{\odot}$) & (M$_{\odot}$) & (M$_{\odot}$) &	($10^{51}$\,erg)\\ 	
\hline
\hline

M25\_Z0.00\_Rot0.0     &	25.0  	&	70069  &  2.01   & 4.94 & 24.99  &    10319  & 195 & 5.58  &  1.70 & 23.29 &  0.001 & 1.0	\\

M25\_Z0.00\_Rot0.2     &	25.0  	&	69877  &  2.02   & 4.94  & 24.99  &    17216  & 57 & 5.42 & 2.00 & 22.99 &  0.001 & 1.0		\\

M25\_Z0.00\_Rot0.4     &	25.0  	&	68882  &  2.07   & 4.94 & 24.96  &    29784  & 32 & 5.87 & 1.80 & 23.16 &  0.001	& 1.0	\\

M25\_Z0.00\_Rot0.6     &	25.0  	&	67069  &  2.16    & 4.93 & 11.94  &    140668  & 1.5 &  5.90 & 2.10 & 9.84 &  0.05 & 1.0	\\

M25\_Z0.00\_Rot0.8     &	25.0  	&	65107  &  2.26    & 4.91 & 11.79 &    175858  & 0.6 & 5.53 & 2.10 & 9.69 &  0.05	& 1.0	\\

M25\_Z0.02\_Rot0.0     &	25.0  	&	3962  &  5.91    & 4.88 & 22.64  &    3623  & 1219 & 5.36 & 1.90 & 22.74 &  0.001 & 1.0 		\\

\hline\hline
\end{tabular}}
\end{center}
%\par
{$^a$Mass at ZAMS.
$^b$Progenitor radius at ZAMS,
$^c$Luminosity at ZAMS,
$^d$Final mass of pre-SN model,
$^e$Pre-SN phase radius,
$^f$Pre-SN phase luminosity,
$^g$Mass of the central remnant in simulation,
$^h$Ejecta mass,
$^i$Amount of synthesised Nickel used in the explosion,
$^j$Explosion energy.}\\

\end{table*} 

\section{Synthetic explosions using {\tt SNEC}}
\label{sec:snec}
Once the models have reached the stage of core-collapse marked by the infall velocity exceeding the specified Iron-core infall velocity, the outputs of {\tt MESA} in appropriate forms are provided as input to {\tt SNEC} \citep[][]{2015ApJ...814...63M}. {\tt SNEC} is a 1-D Lagrangian hydrodynamic code that simulates the synthetic explosions of the stellar models at the stage of the onset of their core collapse. {\tt SNEC} solves the radiation energy transport equations within the flux-limited diffusion approximation to simulate the explosions.

Further, in this work, to simulate the synthetic explosions of the models which have already arrived at the stage of the onset of core collapse, we closely follow the setups of \citet[][]{2019ApJ...877...92O} along with \citet[][] {2021MNRAS.505.2530A, 2022MNRAS.517.1750A} for {\tt SNEC}. However, the major changes are summarized here. 
First, for each model, the innermost mass $M_{\rm c}$ representing the mass of the central remnant is excised before the explosion by assuming that the model will finally collapse to form a neutron star. The central remnant mass is decided by the final mass of the Iron-core when the model has reached the stage of the onset of core collapse. Further, a set of 800 grid cells are used to simulate the synthetic explosion of the model. With 800 grid cells, the light curves and photospheric velocities of the resulting SN from simulations are very well converged in the interested domains of time. The explosion of each model is simulated as {\tt thermal bomb} by adding $E_{\rm exp}$ amount of energy for a duration of 0.1\,s in the inner 0.1\,M$_{\odot}$ section of the model. As discussed in \citet[][]{2015ApJ...814...63M}, {\tt SNEC} lacks nuclear-reaction network, thus, the synthesised amount of Nickel ($^{56}$Ni) in an SN is decided, and fixed by the individual user. For each model, an amount of $^{56}$Ni specified by corresponding $M_{\rm Ni}$ in Table~\ref{tab:MESA_MODELS}, is distributed between the excised central remnant mass ($M_{\rm c}$) cut and the chosen mass coordinate which is close to the outer surface of the selected model. For models with $\Omega$ / $\Omega_{\rm crit}$\,$\leq$\,0.4, the amount of $^{56}$Ni is set to 0.001\,M$_{\odot}$ while the remaining models with heavy rotations and suffering significant mass losses, the amount of $^{56}$Ni is set to 0.05\,M$_{\odot}$. Choosing a slightly greater amount of $M_{\rm Ni}$ for stripped models is followed by 
\citet[][]{2021ApJ...918...89A}. The ejecta mass ($M_{\rm ej}$) for each CCSN is estimated by finding the difference between the pre-SN mass ($M_{\rm Pre-SN}$) and $M_{\rm c}$. The detailed explosion parameters are listed in Table~\ref{tab:MESA_MODELS}.

Finally, the UBVRI -bands light curves generated through synthetic explosions are shown in Figure~\ref{fig:lc}. As shown in this figure, the slow-rotating models (i.e., models M25\_Z0.00\_Rot0.0, M25\_Z0.00\_Rot0.2, M25\_Z0.00\_Rot0.4, and the solar metallicity model M25\_Z0.02\_Rot0.0) retaining a significant amount of their outer H-envelope result into Type II CCSNe while the rapidly rotating models result into CCSNe Type Ib/c. These results are also complementing the results as predicted in the phase diagram of \citet[][]{2012A&A...542A.113Y} (in Figure 12). 
In Figure~\ref{fig:lc}, a few important simulation results of the H-rich Pop III CCSNe (i.e. models with $\Omega$ / $\Omega_{\rm crit}$\,$\leq$\,0.4) are also displayed; First, the peak magnitudes of the shock breakout (SBO) features from these models are much fainter than a typical solar metallicity H-rich CCSN;  second, the absolute magnitudes of the plateau of the H-rich Pop III CCSNe are at least 1.5 magnitudes fainter than the solar metallicity H-rich CCSN, thus the H-rich Pop III CCSNe are pretty faint within the considered limits of $E_{\rm exp}$ and $M_{\rm Ni}$ in this study. The effect of bolometric light curves becoming less luminous as metallicity decreases has been explored in \citet[][]{2009ApJ...703.2205K} and \citet[][]{2018ApJS..234...34P}. The primary cause of this behaviour is associated with the smaller pre-SN radius and less total mass loss as an effect of lower metallicity. However, in their work, they have not calculated the light curves corresponding to Z $=$ 0.00; and the third, surprisingly, although the pre-SN radius of the non-rotating H-rich Pop III model is much smaller than a non-rotating solar metallicity H-rich model, the earlier model shows almost a similar plateau duration. The non-rotating solar model has a larger pre-SN radius compared to the M25\_Z0.00\_Rot0.0 model, but the latter has a more massive H-envelope. From Figure~\ref{fig:mass_fraction}, the M25\_Z0.02\_Rot0.0 model has an H-envelope starting from a mass coordinate, m(M$_{\odot}$)$\sim$8\,M$_{\odot}$ while the M25\_Z0.00\_Rot0.0 has a more-massive H-envelope starting from m(M$_{\odot}$)$\sim$5\,M$_{\odot}$. Thus, the presence of extra Hydrogen could be responsible for the increased plateau duration in the non-rotating Pop III model. Finally, the rapidly rotating H-less Pop III models result into much fainter Type Ib/c SNe. These explosions are fainter because of the less explosion energy (and an M$_{\rm Ni}$ of 0.05\,M$_{\odot}$) considered in our study. With higher explosion energies and more Nickel production, they might result in more luminous SNe or hypernovae \citep[][]{2013ARA&A..51..457N}. 
{\tt SNEC} could also produce the bolometric luminosity light curves and the corresponding photospheric velocity evolutions for all the models as shown in the left and the right panels of Figure~\ref{fig:blc_vel}, respectively. In the left panel, the bolometric light curves of the Pop III CCSNe display a similar behaviour as earlier in the case of UBVRI -bands light curves comparison with the solar metallicity model. In the right panel, as expected, the stripped models display higher photospheric velocities compared to the H-rich models.

\section{Results and Discussion}
\label{sec:results and discussion}
In this work, we have performed the 1-D stellar evolutions of Pop III models up to the stage of the onset core collapse and then simulated their synthetic explosions. Utilising the 1-D simulations performed in this work, we summarize our findings below:
\begin{enumerate}
%    \item Compared to a non-rotating solar metallicity model, the Pop III models arrive on ZAMS at a higher effective temperature. All the rotating models suffer chaotic mass losses, marking dynamic events during their last evolutionary phases.
    
%    \item From Table~\ref{tab:MESA_MODELS}, we find that the pre-SN radii of the Pop III models (either stripped or H-rich) are much smaller than a typical solar Type model. Also, the effective temperatures of Pop III models at various stages are much higher, thus, they are much bluer than a typical solar Type model.

    \item The peak absolute magnitudes of the SBO features of Pop III CCSNe are much smaller than that of a CCSN resulting from a solar Type model with similar ZAMS mass.
    
    \item The H-rich CCSNe from Pop III models are fainter than the H-rich SN resulting from a solar metallicity model. The plateau magnitudes of Pop III star H-rich CCSNe are at least 1.5 magnitudes fainter than the latter. In the earlier epochs, the stripped CCSNe from Pop III models are much fainter than SNe resulting from H-rich Pop III models.
    
    \item One of the most intriguing results from our simulations is that although the pre-SN radius of a non-rotating H-rich Pop III model is much smaller than a non-rotating H-rich solar Type model, both models show nearly similar plateau durations. One of the reasons for the increased plateau duration despite a relatively smaller pre-SN radius in non-rotating Pop III CCSNe could be associated with the increased amount of Hydrogen mass.
    
    \item Among the discussed Pop III models, SN resulting from the non-rotating H-rich model is the brightest. It has a nearly constant absolute magnitude of around -16.5 mag in the V-band for the plateau phase. This would correspond to an apparent magnitude of $\sim$35.5\,mag at a redshift of z = 10 (using the cosmology of a Hubble constant, H$_{0}$ = 73, $\Omega_{\rm M}$ = 0.3, and $\Omega_{\rm vac}$ = 0.7). Currently, no ground- or space-based observatory can go this faint to detect a Pop III CCSN resulting from an individual star, however, with the major advancement in observational technologies having large diameters could possibly detect such events in near future.
    
    \item Thus, through our work, we find that within the considered limits of explosion energies and Nickel masses, these transient events are very faint, making it difficult for them to be detected at high redshifts.
\end{enumerate}

%\section{Discussion and Conclusion}
%\label{sec:discussion}

%%%%%%%%%%%%%%%%%%%%%%%%%%%%%%%%%%%%%%%%%%%%%%%%%%
\section*{Data Availability}
The {\tt inlist} files for {\tt MESA} simulations and {\tt SNEC} input files will be uploaded to {\href{https://zenodo.org/}{\tt zenodo}} publicly. One can download those files to reproduce the simulation results.

\section*{Acknowledgement} 
The authors are highly thankful to the anonymous referee for providing constructive comments for the significant improvement of the Letter. A.A. duly acknowledges the supports furnished by CSIR, India (under the file no. 09/948(0003)/2020-EMR-I). A.A. also sincerely acknowledges useful discussions and suggestions received from {\tt MESA} user community with special thanks to Prof. Francis Timmes. SBP and RG acknowledge the support of ISRO provided under the AstroSat archival Data utilisation program (DS$\_$2B-13013(2)/1/2021-Sec.2).

\bibliographystyle{mnras}
\bibliography{PopIII} %
%%%%%%%%%%%%%%%%%%%%%%%%%%%%%%%%%%%%%%%%%%%%%%%%%%

%%%%%%%%%%%%%%%%% APPENDICES %%%%%%%%%%%%%%%%%%%%%

\appendix

\section{Additional Figures and Tables}
\label{appex1}

\begin{figure*}
%    \centering
    \includegraphics[height=5cm,width=0.33\textwidth,angle=0]{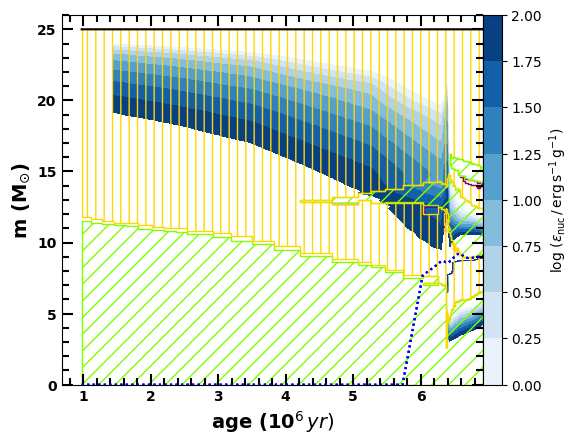}
    \includegraphics[height=5cm,width=0.33\textwidth,angle=0]{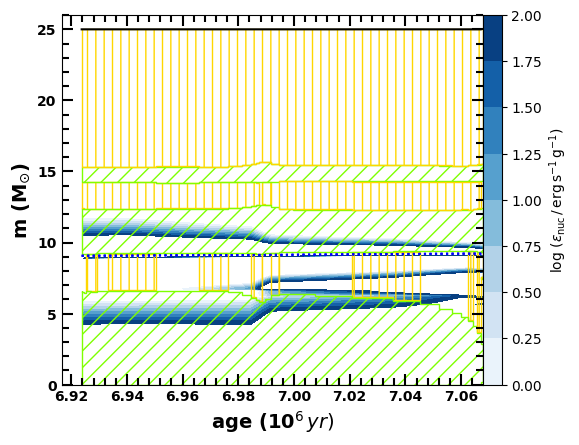}
    \includegraphics[height=5cm,width=0.33\textwidth,angle=0]{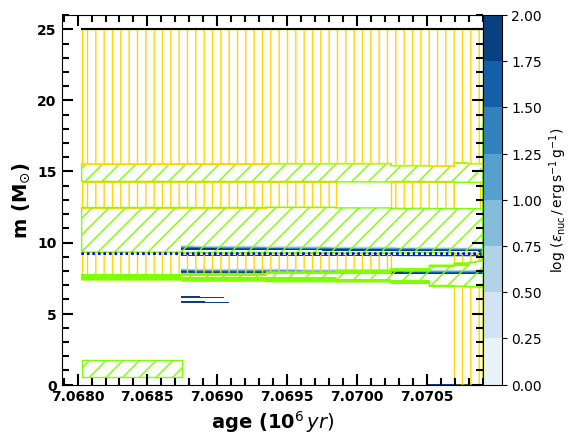}
    \includegraphics[height=5cm,width=0.33\textwidth,angle=0]{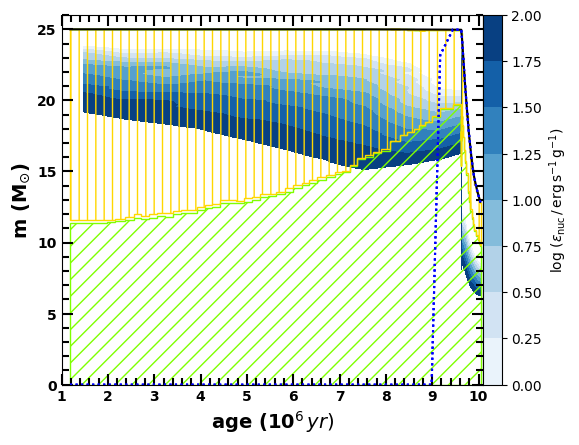}
    \includegraphics[height=5cm,width=0.33\textwidth,angle=0]{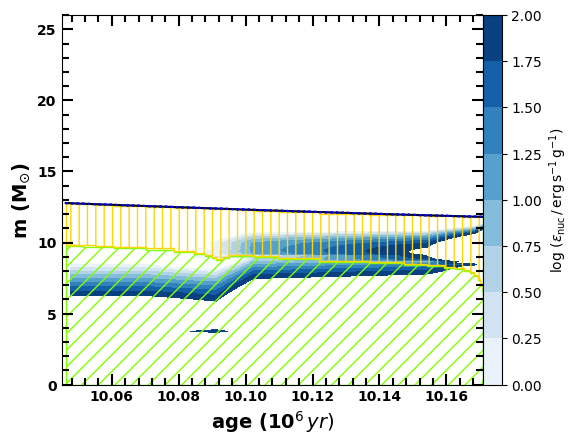}
    \includegraphics[height=5cm,width=0.33\textwidth,angle=0]{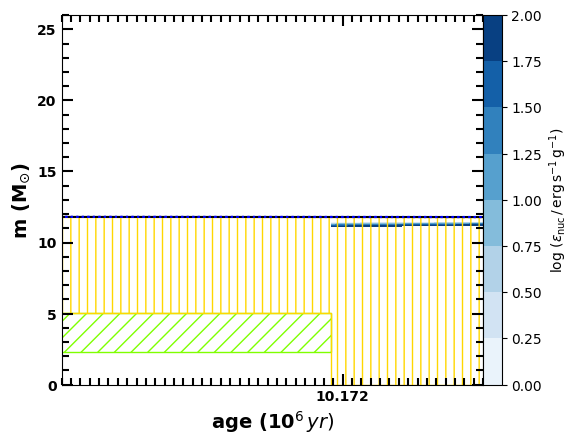}
   \caption {The kippenhahn diagrams of the models M25\_Z0.00\_Rot0.2 (top) and M25\_Z0.00\_Rot0.8 (bottom) for a period between ZAMS to close to the pre-SN stage. Here, the green hatchings indicate the convective regions and the dark-yellow regions mark the stellar interiors where the thermohaline mixing is going on. Also, the logarithm of the specific nuclear energy generation rate ($\epsilon_{\rm nuc}$) inside the stellar interiors is indicated by the blue colour gradients. The rapidly rotating model is significantly stripped.}
    \label{fig:kippenhahn}
\end{figure*}

\begin{figure*}
  \centering
  \includegraphics[height=9cm,width=\textwidth]{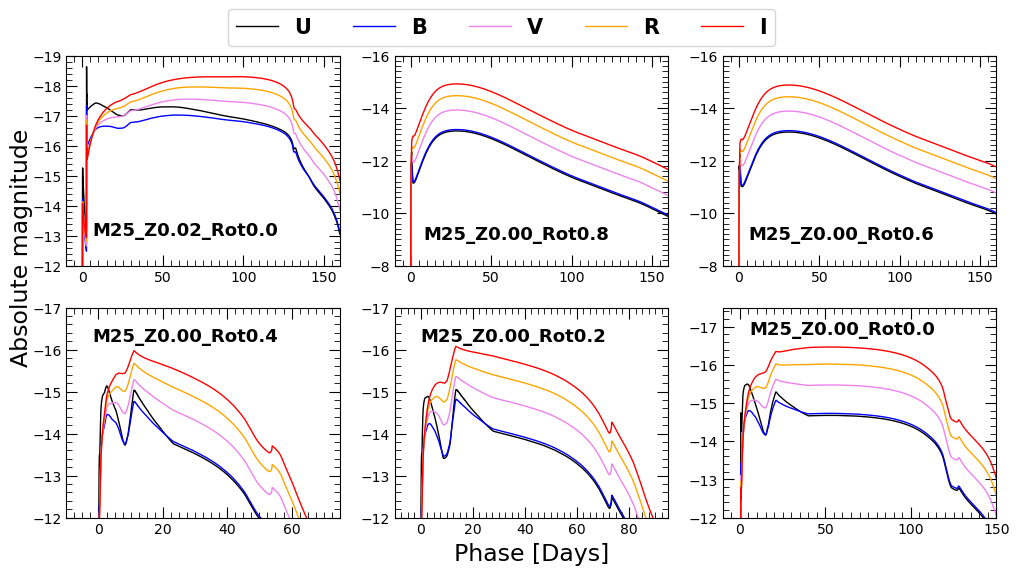}
    \caption{The U, B, V, R, and I band light curves resulting from the synthetic explosions of Pop III models using {\tt SNEC}. The non-rotating and slowly rotating models ($\Omega\,\leq\,0.4\,\Omega_{\rm crit}$) form a class of weak Type II SNe, while the rapidly rotating models ($\Omega\,\ge\,0.6\,\Omega_{\rm crit}$) result into Type Ib/c SNe within the specified limits of explosion energies and Nickel masses. The light curves resulting from the non-rotating, solar metallicity model are also shown for comparison.} 
    \label{fig:lc}
\end{figure*}

\begin{figure*}
  \centering
  \includegraphics[height=10cm,width=\textwidth]{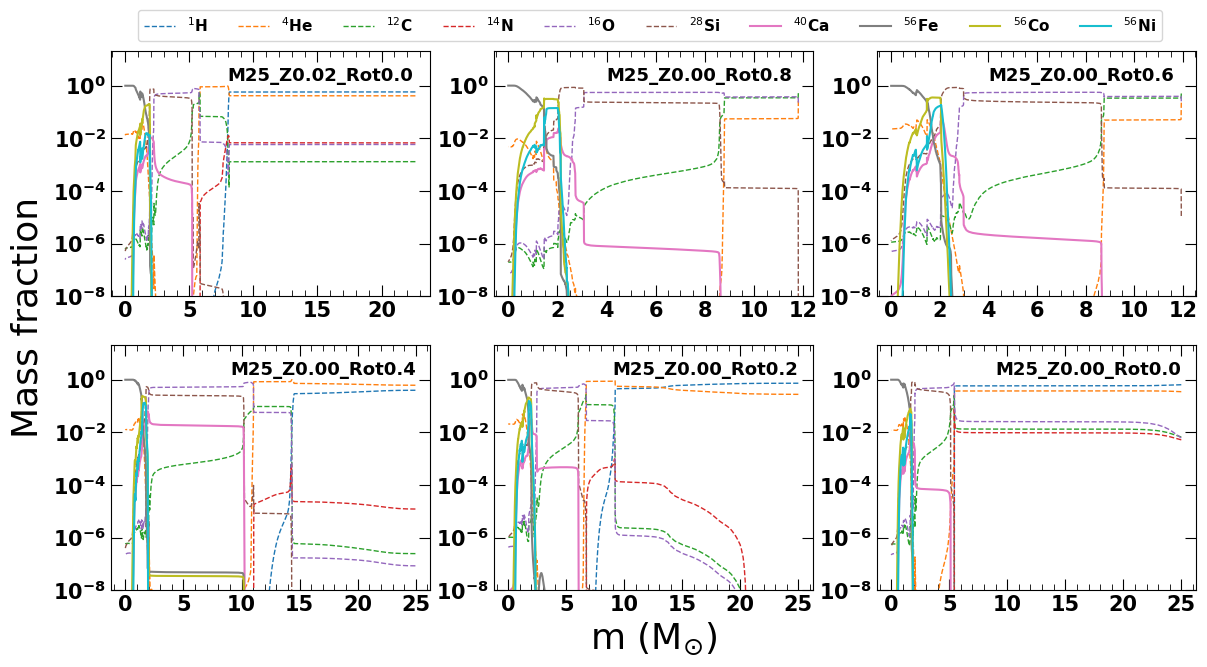}
    \caption{A combined plot showing the mass fractions of various elements for the models in this study at a stage when the models have reached the stage of the onset of core collapse.} 
    \label{fig:mass_fraction}
\end{figure*}
%%%%%%%%%%%%%%%%%%%%%%%%%%%%%%%%%%%%%%%%%%%%%%%%%%

% Don't change these lines
\bsp	% typesetting comment
\label{lastpage}
\end{document}